\documentclass[12pt]{iopart}
\usepackage{iopams}  
\usepackage{epsfig}
\usepackage{float}
\usepackage{wrapfig}
\usepackage{rotating}
\usepackage{cite}
\eqnobysec
\def\eqr#1{equation~(\ref{eq:#1})}
\def\Eqr#1{Equation~(\ref{eq:#1})}
\def\be{\begin{equation}}
\def\ee{\end{equation}}
\def\bea{\begin{eqnarray}}
\def\eea{\end{eqnarray}}
\def\inv{^{-1}}
\def\sigr{\Sigma^{\rm R}}
\def\sigi{\Sigma^{\rm I}}
\def\sigts{\tilde{\Sigma}_{\sigma}}
\def\sigtd{\tilde{\Sigma}_{\downarrow}}
\def\sigtu{\tilde{\Sigma}_{\uparrow}}

\def\sigtur{\tilde{\Sigma}_{\uparrow}^{\rm R}}
\def\sigtsr{\tilde{\Sigma}_{\sigma}^{\rm R}}

\def\sigs{\Sigma_{\sigma}}
\def\sigu{\Sigma_{\uparrow}}
\def\sigd{\Sigma_{\downarrow}}
\def\sigur{\Sigma^{\rm R}_{\uparrow}}

\def\sigui{\Sigma^{\rm I}_{\uparrow}}

\def\sigsr{\Sigma^{\rm R}_{\sigma}}
\def\sigsi{\Sigma^{\rm I}_{\sigma}}
\def\w{\omega}
\def\sgn{{\rm sgn}}

\def\gmb{g\mu_{\rm B}}
\def\sg{{\cal G}}
\def\sa{{\cal A}}
\def\da{\downarrow}
\def\ua{\uparrow}

\def\ra{\rightarrow}
\def\ppm{\Pi^{+-}}
\def\cpm{\chi^{+-}}
\def\pnpm{{^0\Pi^{+-}}}
\def\eg{{\em e.g. }}
\def\ie{{\em i.e. }}
\def\ut{\tilde{U}}
\def\delno{\Delta_0}
\def\pdod{\pi\Delta_0 D}
\def\w{\omega}
\def\wm{{\omega_{\rm m}}}

\def\wp{\omega^\prime}
\def\tp{T^\prime}
\def\wmp{{\omega_{\rm m}^\prime}}
\def\wk{{\omega_{\rm K}}}
\def\tk{{T_{\rm K}}}
\def\tkh{{T_{\rm KH}}}

\def\wt{{\tilde{\omega}}}

\def\tt{\tilde{T}}
\def\im{{\rm Im}}
\def\re{{\rm Re}}
\def\bra{\langle}
\def\ket{\rangle}
\def\half{\frac{_1}{^2}}
\def\PRB{{\em Phys. Rev. B }}

\def\ut{\tilde{U}}

\def\ofw{(\omega)}

\def\ofwo{(\w;0)}
\def\ofwt{(\w;T)}
\def\ofwtil{(\tilde{\omega})}
\def\ofwpt{(\w^+;T)}
\def\ofwpo{(\w^+;0)}
\def\ofot{(0;T)}
\def\oft{(T)}
\def\ofo{(0)}

\def\wct{\tilde{\w}_{\rm c}}

\def\gt{G^{\rm T}}
\def\gst{G^{\rm T}_\sigma}
\def\teo{T\!=\!0}
\def\intall{\int\limits^\infty_{-\infty}}
\def\intp{\int\limits^\infty_0}
\def\fb{f_{\rm B}}
\begin{document}

\title{Finite temperature dynamics of the Anderson model.}

\author{David E. Logan and Nigel L. Dickens }

\address{University of Oxford, Physical and Theoretical Chemistry\\ Laboratory, South Parks Rd, Oxford OX1 3QZ, UK}

\begin{abstract}
The recently introduced local moment approach (LMA) is extended to encompass single-particle dynamics and transport properties of the Anderson impurity model at finite-temperature, $T$.  While applicable to arbitrary interaction strengths, primary emphasis is given to the strongly correlated Kondo regime (characterized by the $\teo$ Kondo scale $\wk$).  In particular the resultant universal scaling behaviour of the single-particle spectrum  $D\ofwt \equiv F( \frac{\w}{\wk}; \frac{T}{\wk})$ within the LMA is obtained in closed form; leading to an analytical description of the thermal destruction of the Kondo resonance on all energy scales.  Transport properties follow directly from a knowledge of $D\ofwt$.  The $T / \wk$-dependence of the resulting resistivity $\rho\oft$, which is found to agree rather well with numerical renormalization group calculations, is shown to be asymptotically exact at high temperatures; to concur well with the Hamann approximation for the s-d model down to $T/\wk \sim 1$, and to cross over smoothly to the Fermi liquid form $\rho\oft - \rho\ofo \propto -(T/\wk)^2$ in the low-temperature limit.  The underlying approach, while naturally approximate, is moreover applicable to a broad range of quantum impurity and related models.
\end{abstract}


\submitto{\JPCM}


\section{Introduction.}
\label{sec:1}

The Anderson impurity model (AIM) \cite{ref:1}, reviewed fulsomely in \cite{ref:2}, remains as topical as ever.  While traditionally the archetype for understanding \cite{ref:2} magnetic impurities in metals, and heavy fermion systems when coherence effects are suppressed by disorder or alloying, study of it has recently acquired further impetus with the advent of quantum dots \cite{ref:3,ref:4}, which form directly tunable mesoscopic realizations of the model.

In the strong coupling domain of large local Coulomb interaction ($U$), the intrinsic low-energy physics of the AIM is of course that of the Kondo effect \cite{ref:2}.  Manifest dynamically in the many-body Kondo resonance arising in the single-particle spectrum $D(\w)$, the effect is characterized by a single low-energy Kondo scale, $\wk$.  In consequence, $D(\w)$ exhibits scaling in terms of $\w/\wk$ alone, with no explicit dependence on bare material parameters. What then is the form of this universal scaling spectrum? That the question itself is obvious does not vitiate the difficulties involved in answering it. Conventional perturbation theory cannot in general handle strong interactions, necessitating the development of new, non-perturbative theoretical approaches; the difficulties being particularly acute when dealing with dynamical and transport properties. And while Fermi liquid theory determines the low-frequency form \cite{ref:2} $D(\w)-D(0) \propto -(|\omega|/\omega_{K})^{2}$, such behaviour is limited to the lowest of frequencies $|\w|/\wk \ll 1$, and otherwise provides no clues to the general form of the Kondo resonance.

We have recently considered this problem \cite{ref:5} via the local moment approach (LMA) \cite{ref:6,ref:7,ref:8,ref:9}. This non-perturbative method, naturally approximate but physically and technically quite simple, has been  developed to handle a broad class of quantum impurity models, as well as lattice-based problems such as the Hubbard or periodic Anderson models within the local framework of dynamical mean-field theory \cite{ref:10,ref:11}. For the normal (metallic host \cite{ref:1}) AIM considered in \cite{ref:5}, the LMA provides a simple analytical description of the Kondo scaling spectrum on all energy scales $\w/\wk$. It shows in particular that while Fermi liquid behaviour is naturally recovered at sufficiently low frequencies, the scaling spectrum is entirely dominated for $|\w|/\wk \gtrsim 1$ by long tails that exhibit a  very slow logarithmic decay; in contrast to either the power-law (Doniach-${\rm \breve{S}unji\acute{c}}$) behaviour previously thought to arise from empirical fits to quantum Monte Carlo \cite{ref:12,ref:13} and numerical renormalization group (NRG) \cite{ref:14,ref:15} calculations, or the simple Lorentzian form \cite{ref:2} inferred \eg from crude extrapolation of low-frequency Fermi liquid behaviour. The resultant theory, while naturally approximate, was found \cite{ref:5} to give very good agreement for essentially all frequencies with recent NRG calculations \cite{ref:15}.

The above comments refer of course to zero temperature ($T$), and the obvious question is whether the LMA can be extended to finite-$T$. If so, then for the strong coupling Kondo regime in particular, the single-particle spectrum $D \equiv D\ofwt$ should scale universally in terms of $\w/\wk$ {\em and}  $T/\wk$, with $\wk$ again the $\teo$ Kondo scale; and the approach then affords an analytical handle on the ($\w$-differential) thermal destruction of the Kondo resonance. In addition, and importantly, knowledge of the $T$-dependence of the single-particle spectrum enables direct access to transport properties \cite{ref:2}, notably the resistivity $\rho(T)$.

It is these issues we consider here, focussing as in \cite{ref:5} on the symmetric AIM, since it is this case whose low-energy physics reduces in strong coupling to the usual Kondo model (with exchange but not potential scattering). The relevant background is introduced in \S2; and in \S3 practical extension of the LMA to finite-$T$ is considered. For finite interaction strengths $\ut = U/\pi\delno$ (with $\delno$ the one-electron hybridization), the LMA captures the spectra on all energy scales, including \eg the non-universal Hubbard satellites. In \S4 a brief discussion is thus first given of the thermal evolution of the spectrum on all energy scales; followed by consideration of spectral scaling as the Kondo limit is approached with increasing $\ut$, where a scaling form $\pdod\ofwt = F\left(\frac{\omega}{\wk}; \frac{T}{\wk}\right)$ is correctly recovered.

It is the scaling regime on which we naturally focus in the remainder of the paper, developing an analytical description thereof in \S5. The asymptotic behaviour of the scaling spectrum in particular, encompassing high-frequencies for all temperatures and {\em vice versa}, is obtained explicitly; and, as for the $\teo$ limit \cite{ref:5}, is found to be largely independent of the details of the LMA. The temperature dependence of the resultant LMA resistivity, which is found to agree rather well with finite-$\ut$ NRG calculations \cite{ref:16}, is considered in \S6; where we show it to be asymptotically exact for $T/\wk \gg 1$ \cite{ref:17}, to agree well with the Hamann formula \cite{ref:18} for the resistivity of the Kondo/s-d model down to $T/\wk \sim 1$, and to cross over to characteristic low-temperature Fermi liquid form \cite{ref:19} $\rho(T)-\rho(0) \propto -(T/\wk)^{2}$ for $T/\wk \ll 1$. Some closing remarks are given in \S7.

\section{Background.}
\label{sec:2}

We begin with requisite background material, starting with the familiar AIM Hamiltonian \cite{ref:1,ref:2}:

\be
\label{eq:ham}
\hat{H} = \sum\limits_{\bi{k}, \sigma} \epsilon_{\bi{k}} \hat{n}_{\bi{k} \sigma} + \sum\limits_\sigma \left(\epsilon_i + \frac{U}{2} \hat{n}_{i -\sigma} \right) \hat{n}_{i\sigma} + \sum\limits_{\bi{k},\sigma}V_{i\bi{k}}\left( c_{i\sigma}^\dagger c_{\bi{k}\sigma} + \rm{h.c.} \right)
\ee

\noindent The first term describes the non-interacting host band with dispersion $\epsilon_\bi{k}$; while the second refers to the correlated impurity with on-site interaction $U$, and site-energy $\epsilon_i = -\frac{U}{2}$ for the particle-hole (p-h) symmetric AIM (for which the impurity charge $n_i = \sum_\sigma\bra \hat{n}_{i\sigma} \ket = 1$  for all  $U$ and $T$).  The final term in \eqr{ham} is the host-impurity coupling.

In considering single-particle dynamics at finite-$T$, we focus naturally on the retarded impurity Green function $G\ofwt$ ($\leftrightarrow G(t;T) = -\rmi\theta(t)\bra\{c_{i\sigma}(t),c_{i\sigma}^\dagger\}\ket$), with the single-particle spectrum $D\ofwt = -\frac{1}{\pi}\im~G\ofwt$.  In the non-interacting limit $U=0$, $G\ofwt$ reduces trivially to $g(\w) = [\w^+ - \Delta(\w)]\inv$, with $\w^+ = \w + \rmi 0^+$.  Here $\Delta(\w) = \sum_\bi{k} V_{i\bi{k}}^2 [\w^+ - \epsilon_\bi{k}]\inv$ ( $=\Delta_{\rm{R}}(\w) - \rmi \Delta_{\rm{I}}(\w)$) is the one-electron hybridization function such that $\Delta_{\rm{I}}\ofw = \pi\sum_\bi{k}V_{i\bi{k}}^2 \delta(\w - \epsilon_\bi{k}) = \Delta_{\rm{I}}(-\w)$; and the usual hybridization strength is defined by $\delno = \Delta_{\rm{I}}(\w=0) \propto \rho_{\rm{host}} (\w = 0)$ (with $\w = 0$ the Fermi level).  From a knowledge of $D\ofwt$ generally, and hence the transport time $\tau\ofwt \propto [\delno D\ofwt]\inv$, impurity contributions to transport properties follow via the transport integrals \cite{ref:2,ref:16,ref:20}.  The resistivity $\rho(T)$ is given in particular by

\be
\label{eq:rhodef}
\left[ \frac{\rho(T)}{\rho(0)} \right]\inv = \int\limits^\infty_{-\infty} \rmd \w~\frac{1}{\pdod \ofwt} \left( - \frac{\partial f \ofwt}{\partial \w} \right)
\ee

\noindent with $f\ofwt$ the Fermi function; it will be considered explicitly in \S6.  The thermal conductivity, Hall coefficient etc. are likewise expressible \cite{ref:2,ref:16,ref:20} as appropriate moments of $\tau^l\ofwt~ \partial f\ofwt / \partial \w$.

The essential strategy behind the LMA \cite{ref:5,ref:6,ref:7,ref:8,ref:9} is threefold.  (i) Local moments, regarded as the first effect of electron interactions, are introduced explicitly from the outset \cite{ref:1}.  Despite the stark deficiencies of static mean-field theory (MF) by itself (\ie unrestricted Hartree-Fock), it may nevertheless be used as a starting point for a genuine many-body approach encompassing the correlated electron dynamics that are the essence of the Kondo effect.  (ii)  To this end the LMA invokes a two-self-energy description that is a natural consequence of the underlying {\em two} degenerate, broken symmetry MF states; introducing non-trivial dynamics into the associated self-energies $\Sigma_\sigma \ofwt$ via coupling of single-particle excitations to low-energy transverse spin fluctuations.  (iii) The final key element behind the LMA for $\teo$, likewise discussed further below, is that of symmetry restoration: self-consistent restoration of the broken symmetry endemic at MF level.  If symmetry can be restored (as is always the case for the metallic AIM considered here), then self-consistent imposition of it reveals the finite timescale $\tau \sim h/\wm$ on which it occurs; the associated energy $\wm \propto \wk$ being the Kondo scale, whose origin within the LMA thus stems directly from symmetry restoration.

The impurity Green function is expressed formally as \cite{ref:5,ref:6,ref:7,ref:8,ref:9}

\numparts
\be
\label{eq:gdef1}
G\ofwt = \half \sum\limits_\sigma G_\sigma \ofwt
\ee

\noindent where

\be
\label{eq:gdef2}
G_\sigma \ofwt = [\w^+ - \Delta(\w) - \sigts \ofwpt]\inv
\ee
\endnumparts

\noindent (and $\sigma = \ua/\da$ or $+/-$).  The corresponding self-energies $\sigts \ofwpt$ are separated as

\be
\label{eq:sigtdef}
\sigts \ofwpt = -\frac{\sigma}{2} U |\mu| + \sigs \ofwpt
\ee

\noindent into (i) a purely static Fock contribution (with local moment $|\mu|$) that alone would survive at MF level (the trivial Hartree piece of $\frac{U}{2} n_i = \frac{U}{2}$ is canceled precisely by $\epsilon_i = -\frac{U}{2}$); and (ii) an $\w$-dependent contribution $\sigs \ofwpt$ containing in particular the spin dynamics that dominate the low-energy physics of the problem.  The $G_\sigma \ofwt$ may be cast equivalently as

\be
\label{eq:gfunc}
G_\sigma \ofwt = [{\cal G}_\sigma (\w)\inv - \sigs \ofwpt ]\inv
\ee

\noindent in terms of the MF propagator

\be
\label{eq:scriptg}
{\cal G}_\sigma (\w) = [\w^+ +\frac{\sigma}{2} U|\mu| - \Delta (\w)]\inv,
\ee

\noindent and with $\sigs \equiv \sigs[\{{\cal G}_\sigma\}]$ a functional of $\cal{G}_\ua$, $\cal{G}_\da$.  [Note that the ${\cal G}_\sigma(\w)$ here refer to, and $\sigs$ and $G_\sigma$ are thus constructed from, (either) one of the underlying degenerate MF states; and that the rotational invariance of the problem is correctly preserved by the `spin-sum' \eqr{gdef1} as shown directly in \cite{ref:6}.]  Separating the retarded self-energies as

\be
\label{eq:sigparts}
\sigs \ofwpt = \sigsr \ofwpt - \rmi\sigsi \ofwpt
\ee

\noindent analyticity demands that $\sigsi \ofwpt$ be positive semi-definite for all $\w$ and $T$; while p-h symmetry implies

\be
\label{eq:phsigsym}
\sigsi \ofwpt = \Sigma_{-\sigma}^{\rm I} (-\w^+;T) \hspace{14mm} \sigsr \ofwpt = -\Sigma_{-\sigma}^{\rm R} (-\w^+;T),
\ee

\noindent such that ($-\frac{1}{\pi} \im~G_\sigma \ofwt =$) $D_\sigma \ofwt = D_{-\sigma} (-\w;T)$ and hence $D \ofwt = D (-\w;T)$ for the impurity spectrum.

For $\teo$ it is of course more natural/convenient to work with time-ordered self-energies, but these are simply related to their retarded counterparts.  We denote the $\teo$, $t$-ordered Green functions by $\gt\ofw$ and $\gst \ofw$ (for which equations~(2.3) hold with $\w^+ = \w + \rmi0^+\sgn\ofw$); and the corresponding self energies by $\sigs\ofw$, such that $\sigs \ofw = \sigsr \ofw - \rmi~\sgn\ofw \sigsi \ofw$ with $\sigsi \ofw \geq 0 ~ \forall ~ \w$.  The $t$-ordered and $\teo$ retarded Green functions are then related simply by $\re~G_\sigma(\w; \teo) = \re~\gst \ofw$ and $\im~ G_\sigma (\w;\teo) = \sgn\ofw \im~\gst \ofw$; and likewise:

\be
\label{eq:t2retsigs}
\sigsr (\w^+;\teo) = \sigsr \ofw  \hspace{14mm} \sigsi(\w^+;\teo) = \sigsi \ofw
\ee

\begin{wrapfigure}[14]{r}{65mm}
\centering\epsfig{file=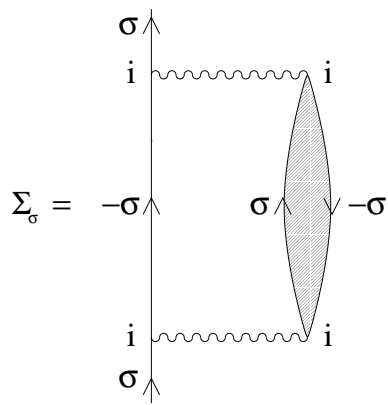,width=50mm,angle=0}
{\protect\caption{{Principal contribution to the LMA $\Sigma_\sigma(\omega)$, see text.  Wavy lines denote $U$.}}}
\label{fig:diag}
\end{wrapfigure}

\noindent  The LMA includes in $\sigs \ofw$  the non-perturbative class of diagrams shown in figure~1, that embody dynamical coupling of single-particle excitations to low-energy transverse spin fluctuations.  These capture the spin-flip scattering essential to describe the strong coupling Kondo regime for $\ut = U/\pi \delno \gg 1$; while in weak coupling they ensure the LMA is perturbatively exact to/including second order in $U$ \cite{ref:6}.  Other classes of diagrams may also be included \cite{ref:6}, but are of minor importance; retention of the dynamical spin-flip scattering processes is however essential, and it is these on which we focus.  Figure~1 translates to

\vspace{5mm}

\be
\label{eq:diag}
\sigs \ofw = U^2 \int\limits^\infty_{-\infty}\frac{\rmd\w_1}{2\pi\rmi}~{\cal G}_{-\sigma}(\w - \w_1) \Pi^{-\sigma\sigma}(\w_1)
\ee

\noindent where $\Pi^{-\sigma\sigma} (t) = \rmi \bra T(S_i^{-\sigma}(t) S_i^{\sigma})\ket$ is the ($t$-ordered, $\teo$) transverse spin polarization propagator (shown hatched in figure~1).  It is given at the simplest level by an RPA-like particle-hole ladder sum in the transverse spin channel; viz

\be
\label{eq:pisum}
\Pi^{-\sigma\sigma}\ofw = {^0\Pi}^{-\sigma\sigma}\ofw\left[ 1 - U^0\Pi^{-\sigma\sigma}\ofw\right]\inv
\ee

\noindent with $^0\Pi^{-\sigma\sigma}\ofw$ the bare p-h bubble, itself expressed in terms of the broken symmetry MF propagators $\{ {\cal G}_\sigma\}$ (solid lines in figure~1).

The final step in the LMA for $\teo$ \cite{ref:5,ref:6,ref:7,ref:8,ref:9} is self-consistent imposition of symmetry restoration (SR), embodied in the condition $\sigtu(\w=0) = \sigtd(\w=0)$ at the Fermi level $\w=0$; and hence for the p-h symmetric AIM (using equations~(2.4,8,9)) by

\be
\label{eq:pinning}
\left( \sigur (0^+;\teo) \equiv \right)\quad \sigur (0) = \half U |\mu|.
\ee

\noindent For given $\ut$, \eqr{pinning} amounts in practice to a self-consistency equation for the local moment $|\mu|$ \cite{ref:6}.  In physical terms, its consequences in general are threefold \cite{ref:6,ref:7}.  (i) Satisfaction of the SR condition guarantees that the low-energy properties of the system amount to a renormalization of the non-interacting limit, \ie that the system is a Fermi liquid with well defined quasiparticle behaviour.  (ii) Most importantly, solution of \eqr{pinning} generates a low-energy spin-flip scale $\wm$ that sets the finite timescale $\tau \sim h / \wm$ for restoration of the broken symmetry inherent at crude MF level (stemming in effect from dynamical tunneling between the degenerate MF minima).  Manifest in particular as a strong resonance in $\im\ppm \ofw$ centred (by definition of $\wm$) on $\w = \wm$, this is the Kondo scale; with $\wm \propto \exp (-\pi U / 8\delno)$ in strong coupling.  (iii)  If the SR condition \eqr{pinning} cannot be satisfied then a doubly degenerate local moment phase results \cite{ref:7}:  the spin-flip scale $\wm$ = 0 (as characteristic of the locally degenerate state) reflecting the fact that the broken symmetry/degeneracy cannot be restored ($\tau = \infty$).  This situation does not arise in the metallic AIM considered here, where SR is satisfied for all finite $\ut = U / \pi \delno$.  But it is the self-consistent possibility of such inherent in \eqr{pinning} that enables the LMA to access the quantum phase transition from a Fermi liquid to a local moment state in problems such as the soft-gap AIM \cite{ref:7,ref:15} where the non-Fermi liquid local moment state arises.

Before turning to the LMA for finite-$T$ we note (for use in \S 5 below) that the conventional single self-energy $\Sigma \ofwpt$, defined by $G\ofwt = [\w^+ - \Delta\ofw - \Sigma \ofwpt ]\inv$, is readily obtained as a byproduct of the two-self-energy description intrinsic to the LMA.  Using equations~(2.3) it is given for arbitrary $T$ by

\be
\label{eq:fullsig}
\Sigma \ofwpt = \frac{\frac{1}{2}\left[ \sigtu \ofwpt + \sigtd \ofwpt \right] - g \ofw \sigtu \ofwpt \sigtd \ofwpt     }{ 1 - \half g\ofw \left[ \sigtu \ofwpt + \sigtd \ofwpt \right]     }
\ee

\noindent with $g\ofw = [\w^+ - \Delta \ofw ]\inv $ the $U=0$ propagator.

\section{Finite-$T$ LMA.} 

To extend the LMA to finite temperature we first obtain, and then consider the rather transparent physical content of, the $\teo$ retarded self-energy $\sigs (\w^+; \teo)$.  This follows, using equations~(2.7,9), from its $t$-ordered counterpart $\sigs \ofw$; itself given explicitly by \eqr{diag}, where the transverse spin polarization propagator therein satisfies the Hilbert transform:

\be
\label{eq:pihilb}
\Pi^{-\sigma\sigma} (z) = \int\limits^\infty_{-\infty} {\frac{\rmd \w_1}{\pi}} ~ {\frac{\chi^{-\sigma\sigma} (\w_1;\teo)~ \sgn(\w_1)}{\w_1 - z}}
\ee

\noindent Here, $z= \w + \rmi 0^+~ \sgn(\w)$ for the $t$-ordered polarization propagator $\Pi^{-\sigma\sigma} (\w)$ contained in \eqr{pihilb} (while $z= \w + \rmi 0^+~$ [$=\w^+$] for the corresponding $\teo$ retarded propagator $\Pi^{-\sigma\sigma} (\w^+;\teo)$); and $\chi^{-\sigma\sigma} (\w;\teo) = \im\Pi^{-\sigma\sigma}(\w)$ ($= \sgn(\w)\im\Pi^{-\sigma\sigma}(\w^+;\teo)$) is the corresponding spectral density of transverse spin excitations, such that $\chi^{-\sigma\sigma} (\w;\teo) \geq 0$ and $\chi^{-\sigma\sigma} (\w;\teo) = \chi^{\sigma-\sigma} (-\w;\teo)$ \cite{ref:6}.  Using \eqr{pihilb} in \eqr{diag} leads straightforwardly to

\be
\label{eq:lmasigs}
\hspace{-12mm}\sigs \ofw = U^2\! \int\limits^\infty_{-\infty}\! \frac{\rmd\w_1}{\pi}\!\left[ \theta(\w_1) \sg^+_{-\sigma}(\w\! -\! \w_1) + \theta(-\w_1) \sg^-_{-\sigma} (\w\! -\! \w_1)  \right]\chi^{-\sigma\sigma} (\w_1;T\!=\!0) 
\ee

\noindent where

\be
\label{eq:lmascriptg}
\sg^\pm_\sigma \ofw = \int\limits^\infty_{-\infty} \rmd \w_1 ~ \frac{D_\sigma^0(\w_1)\theta(\pm\w_1)}{\w - \w_1 \pm \rmi0^+}
\ee

\noindent denote the one-sided transforms of the ($t$-ordered) MF propagator in terms of its spectral density $D_\sigma^0\ofw$; and $\theta(x)$ is the unit step function.  The $\teo$ retarded self-energy then follows directly using \eqr{t2retsigs}, viz

\bea
\label{eq:lmaretsigt0}
\nonumber\hspace{-25mm}\sigs(\w^+;T\!=\!0) = U^2 \int\limits^\infty_{-\infty}\frac{\rmd\w_1}{\pi}  \int\limits^\infty_{-\infty} \rmd\w_2~ \\
\left[ \theta(\w_1)\theta(\w_2) + \theta(-\w_1)\theta(-\w_2)  \right] \frac{D_{-\sigma}^0(\w_2)}{\w - \w_1 -\w_2 + \rmi 0^+} \chi^{-\sigma\sigma}(\w_1; T\!=\!0)
\eea

\noindent as sought.

The natural extension of the LMA to finite-$T$ may now be deduced by considering the physical processes to which the two terms in \eqr{lmaretsigt0} for $\teo$ correspond.  We focus explicitly on $\sigd$ in the following; with $f\ofwt = [\exp(\beta\w) + 1]\inv$ the Fermi function, such that $f\ofwo = \theta (-\w)$.  Consider the first term in \eqr{lmaretsigt0} for $\sigd(\w^+;\teo)$, in $\theta(\w_1)\theta(\w_2)$ (noting that $\cpm(\w_1;T)$ for $\w_1 >  0$ corresponds to first flipping the impurity spin from $\ua$ to $\da$).  Physically, this corresponds to processes in which (for $\sigd$) a $\da$-spin electron is first added to an $\ua$-spin occupied impurity, and the originally present $\ua$-spin then hops off the impurity/site into the conduction band.  The latter generates an on-site $\ua \ra \da$ spin-flip, which is a hard core boson in the sense that at most one spin-flip can be created on the impurity; and the ($\teo$) probability with which the $\ua$-spin can be added to/hop into (empty) conduction band states is $\theta(\w_2) = 1 - f(\w_2;\teo)$.  Appropriate extension of this contribution to finite-T is then achieved simply by replacing $\theta(\w_1)\theta(\w_2)$ with $\theta(\w_1)[1-f(\w_2;T)]$ in \eqr{lmaretsigt0} for $\sigd$; with $1-f(\w_2;T)$ now the finite-T probability with which the original $\ua$ spin can hop into conduction band states, and to which process creation of the on-site spin-flip is again slaved.

A directly analogous situation pertains to the second term in \eqr{lmaretsigt0} for $\sigd$, in $\theta(-\w_1)\theta(-\w_2)$ (noting that $\cpm(\w_1; T)$ for $\w_1 < 0$ corresponds to first flipping the impurity spin from $\da$ to $\ua$).  This corresponds physically to processes in which (for $\sigd$) an originally present $\da$-spin on the impurity is first removed, and an $\ua$-spin electron then hops from the conduction band onto the impurity.  The latter process generates the on-site $\da \ra \ua$ spin-flip; and the ($\teo$) probability with which the $\ua$-spin electron may be removed from the (occupied) conduction band states is $\theta(-\w_2) = f(\w_2;\teo)$.  Extension of this contribution to finite-T is then achieved by replacing $\theta(-\w_1)\theta(-\w_2)$ with $\theta(-\w_1)f(\w_2;T)$ in \eqr{lmaretsigt0} for $\sigd$; with $f(\w_2;T)$ the finite-$T$ probability with which an $\ua$-spin electron can be removed from the conduction band to hop onto the impurity, thereby generating the on-site spin-flip.

The finite-$T$ LMA we consider is thus given from the above by

\numparts
\be
\label{eq:retsig1}
\hspace{-10mm}\sigd \ofwpt = U^2 \intall \frac{\rmd\w_1}{\pi} \intall \rmd\w_2 ~ \cpm(\w_1;T)~\frac{D^0_\ua (\w_2)}{\w - \w_1 - \w_2 + \rmi 0^+}~ g(\w_1;\w_2)
\ee

\noindent with

\be
\label{eq:retsig2}
g(\w_1;\w_2) = \theta(\w_1)[1 - f(\w_2;T)] + \theta(-\w_1)f(\w_2;T).
\ee
\endnumparts

\noindent In \eqr{retsig1} we have also naturally included the $T$-dependence of $\cpm\ofwt \geq 0$; where again $\cpm\ofwt = \sgn\ofw \im~\ppm\ofwpt$, with $\ppm\ofwpt$ now the finite-$T$ polarization propagator (for which the Hilbert transform \eqr{pihilb} is again applicable, with $z = \w^+$).  $\sigu \ofwpt$ likewise follows from equation~(3.5) by inverting all spins ($\ua \leftrightarrow \da$, $+ \leftrightarrow -$) or, in the symmetric case of interest, from p-h symmetry \eqr{phsigsym}.

We note that equation~(3.5) is not what arises if one attempts to obtain the appropriate LMA $\sigs \ofwpt$ via a conventional analytical continuation of the corresponding imaginary time $\sigs (\rmi\w_n)$.  In that case it is readily shown that the resultant $\sigd \ofwpt$ is indeed of form \eqr{retsig1}, but with $g(\w_1;\w_2)$ replaced by $\tilde{g}(\w_1;\w_2) = -\sgn(\w_1)[\fb(-\w_1;T) + f(\w_2;T)]$ where $\fb \ofw = [\exp (\beta\w) - 1]\inv$ is the Bose function.  This corresponds physically to treating the impurity spin-flips as {\em free} bosons, with thermal statistics that are entirely divorced from those of the fermions.  In generating multiple thermally created spin-flips, it thus in effect violates the hard core condition for the bosonic spin-flips -- whose probability of creation is dictated by the ($T$-dependent) probability with which fermions can hop to/from the impurity from/to the conduction band, as embodied in equation~(3.5).  The hard core constraint can however be recovered simply by replacing the free Bose function $\fb (-\w_1;T)$ in $\tilde{g}(\w_1;\w_2)$ by its $\teo$ limit $\fb(-\w_1;0) = -\theta(\w_1)$: in that case $\tilde{g}(\w_1;\w_2)$ reduces precisely to $g(\w_1;\w_2)$, and equation~(3.5) for $\sigd\ofwpt$ is recovered.  We do not know what {\em additional} diagrams may be required to recover hard core behaviour within the framework of a conventional analytical continuation; but believe that the physical arguments given above, together with the results of the following sections, attest to the essential validity of equation~(3.5) as used in practice for the LMA $\sigs \ofwpt$.

The finite-$T$ LMA is readily implemented.  The retarded $G\ofwt$ (and hence $D \ofwt = -\frac{1}{\pi}\im~G\ofwt$) is given from equation~(2.3), with the self-energies $\sigts\ofwpt$ from \eqr{sigtdef} and $\sigs\ofwpt$ from equation~(3.5).  The finite-$T$ retarded polarization propagator entering \eqr{retsig1} (such that $\chi^{\sigma-\sigma}\ofwt = \sgn\ofw \im\Pi^{\sigma-\sigma}\ofwpt$) is calculated in practice for finite $\ut$ at the level of the RPA-like p-h ladder sum, obtained by straightforward analytical continuation of the imaginary time $\ppm(\rmi\w_{\rm m}) = \pnpm (\rmi\w_{\rm m}) [ 1 - U \pnpm (\rmi\w_{\rm m})]\inv$ (cf \eqr{pisum}).  Finally, it is also straightforward to include a $T$-dependence for the local moment $|\mu| = |\mu(T)|$ entering the MF propagators (\eqr{scriptg}) and the LMA self-energies (\eqr{sigtdef}).  This may be encompassed via $|\mu(T)| = |\mu(0)| + \delta|\mu(T)|$, with $|\mu(0)|$ the $\teo$ moment required to satisfy the symmetry restoration condition \eqr{pinning}; and with $\delta|\mu(T)|$ calculated in practice at MF level, viz $\delta|\mu(T)| = \int^\infty_{-\infty}\rmd\w~[D^0_\ua(\w) - D^0_\da(\w)][f\ofwt - \theta(-\w)]$ with $D^0_\sigma\ofw = \delno \pi\inv / [(\w + \frac{\sigma}{2}U|\mu(0)|)^2 + \delno^2]$ the MF spectral density. We find however that the resultant $T$-dependence is negligible for essentially all $\ut$ (provided one is not concerned with temperatures on the order of $U$), and we thus omit it from the results shown explicitly in \S's 4ff.

\section{Results.}

The primary interest in single-particle dynamics resides of course in thermal destruction of the low-energy Kondo resonance, particularly in the strong coupling Kondo/spin-fluctuation regime where the AIM maps onto the Kondo or s-d model.  That is our main focus, and is pursued in  the following sections.  First however, we consider the issue of spectral scaling as the Kondo limit is approached with increasing $\ut$; preceded by a brief discussion of the thermal evolution of the spectrum on all energy scales.  The following results, obtained from the finite-$T$ LMA outlined above, refer explicitly to the usual wide-band AIM for which $\Delta_{\rm I} (\w) = \delno ~ \forall ~ \w$.  Here and throughout, the Kondo scale $\wk$ is defined as the half-width at half maximum  of the $\teo$ single-particle spectrum $D\ofwo$; and $\wk \propto \wm$ \cite{ref:5,ref:6}, with the spin-flip scale $\wm$ determined (as above) via symmetry restoration, and exponentially small in strong coupling.

\begin{figure}
\centering\epsfig{file=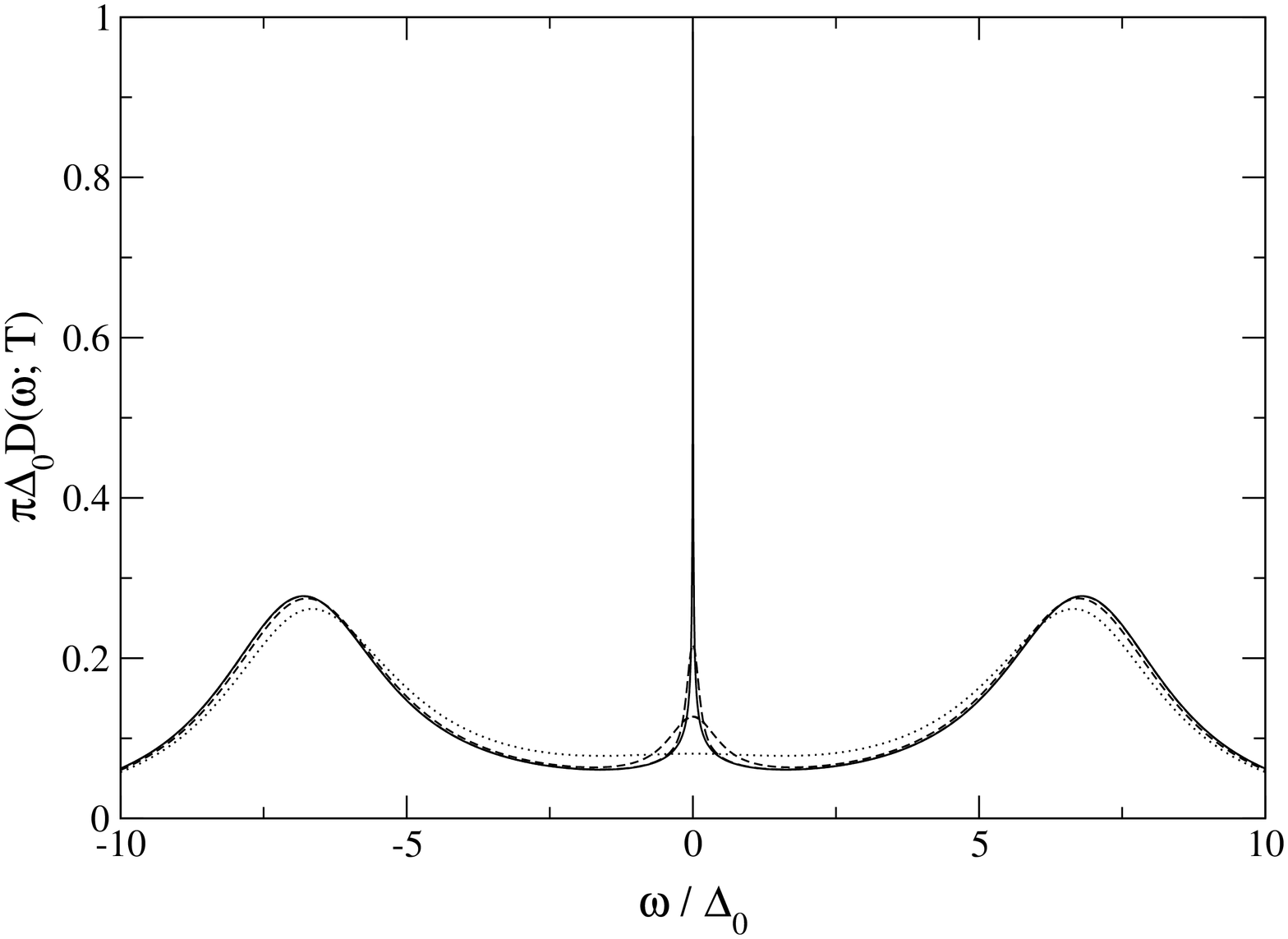,width=100mm,angle=270}
\vskip-5mm
\protect\caption{$\pdod \ofwt$ vs $\w / \delno$ for $\ut = 4$ and $T/\wk=0.1$ (solid line), 10 (long dash), 40 (short dash) and 160 (dotted line).}
\label{fig:2}
\end{figure}

Figure~2 shows the resultant spectrum, $\pdod \ofwt$ vs $\w / \delno$; for a fixed interaction strength $\ut = U / \pi\delno = 4$, and for temperatures $T/\wk = 0.1$, $10$, $40$ and $160$, the intent being an all scales overview of spectral evolution.  The dominant effect of increasing temperature is naturally to erode the many-body Kondo resonance (as later considered in detail).  This process occurs initially `on the spot', with no effect upon spectral features on the non-universal energy scales $\w \sim \delno$ or $U$ \cite{ref:2,ref:16}; as evident in figure~2 for $T/\wk = 0.1$ and $10$, where \eg the high-energy Hubbard satellites retain their $\teo$ form (being centred to high accuracy for the wide-band AIM on $|\w| = \frac{U}{2}\left( 1 + \frac{4\delno}{\pi U}\right)$ \cite{ref:6} ).  By contrast, for non-universal temperature scales on the order $T \sim \delno$, spectral redistribution occurs on all energy scales including the high-energy satellites.  This effect, likewise known to arise from NRG calculations \cite{ref:16}, is evident in figure~2 for $T/\wk = 40$ and $160$ ($T/\delno \sim \frac{1}{4}$ and $1$ respectively).  

We turn now to the question of spectral scaling.  In the strong coupling spin-fluctuation regime, the low-energy physics of the AIM depends solely upon the Kondo scale; and the $\teo$ Kondo resonance hence exhibits universal scaling in terms of $\w / \wk$ alone, with no explicit dependence on the bare material parameters $U$ or $\delno$.  The LMA for $\teo$ leads correctly to such scaling behaviour with increasing $\ut$ \cite{ref:5,ref:6}, and the resultant scaling spectrum gives very good agreement with NRG calculations \cite{ref:5,ref:15}. Scaling behaviour should likewise arise in strong coupling for $T \ne 0$, but with the scaling spectrum $\pdod \ofwt \equiv F(\w / \wk; T/\wk)$ now dependent upon both $\w / \wk$ and $T / \wk$.

\begin{figure}
\centering\epsfig{file=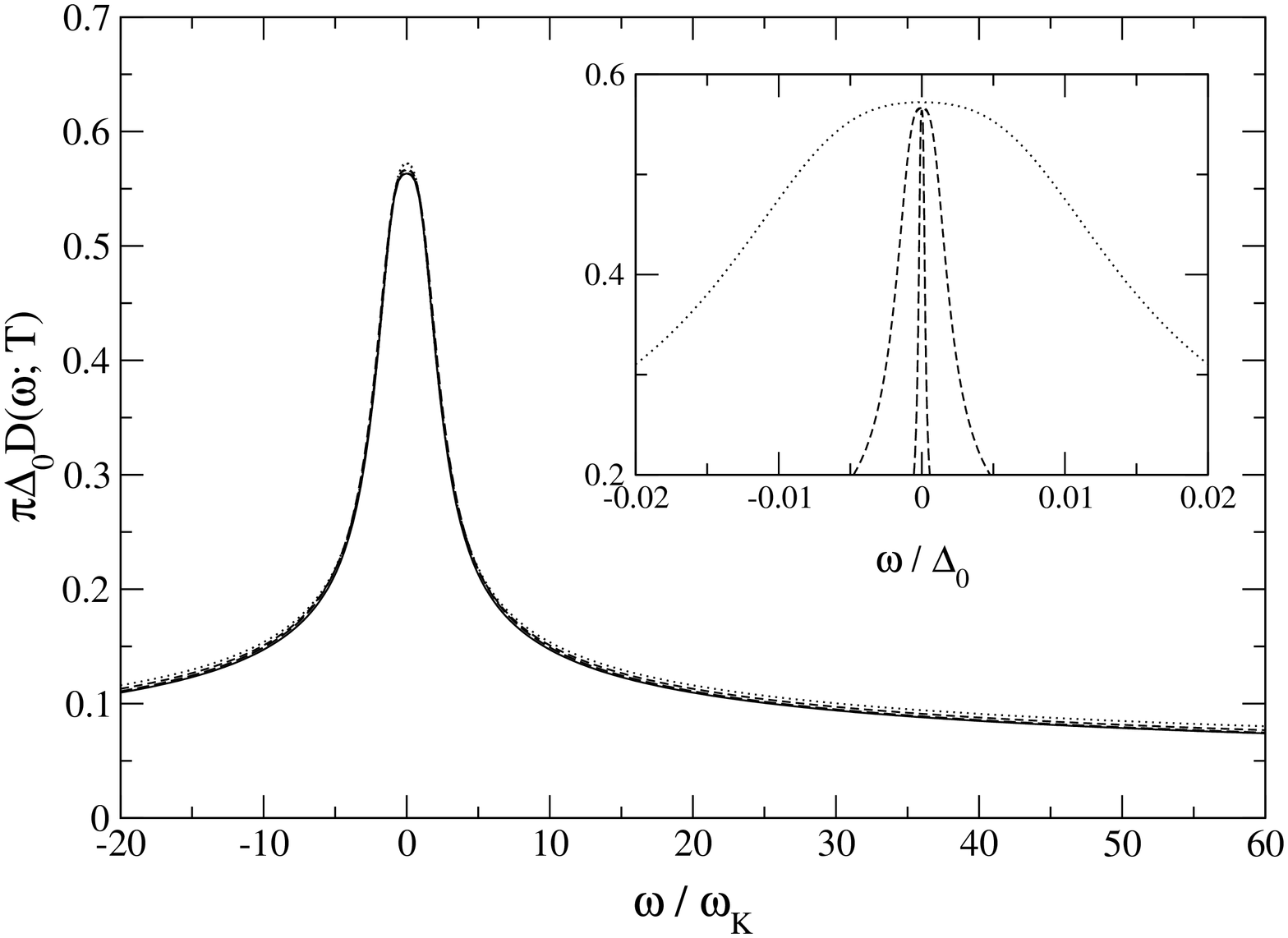,width=100mm,angle=270}
\vskip-5mm
\protect\caption{$\pdod \ofwt$ vs $\w / \wk$ for fixed $T/\wk =1$, and $\ut = 4$ (dotted line), 6 (short dash) and 8 (long dash).  The approach to the Kondo limit scaling spectrum (solid line, see \S5) is evident.  {\bf Inset}: Corresponding spectra on an absolute scale, vs $\w / \delno$.}
\label{fig:3}
\end{figure}

\begin{figure}
\centering\epsfig{file=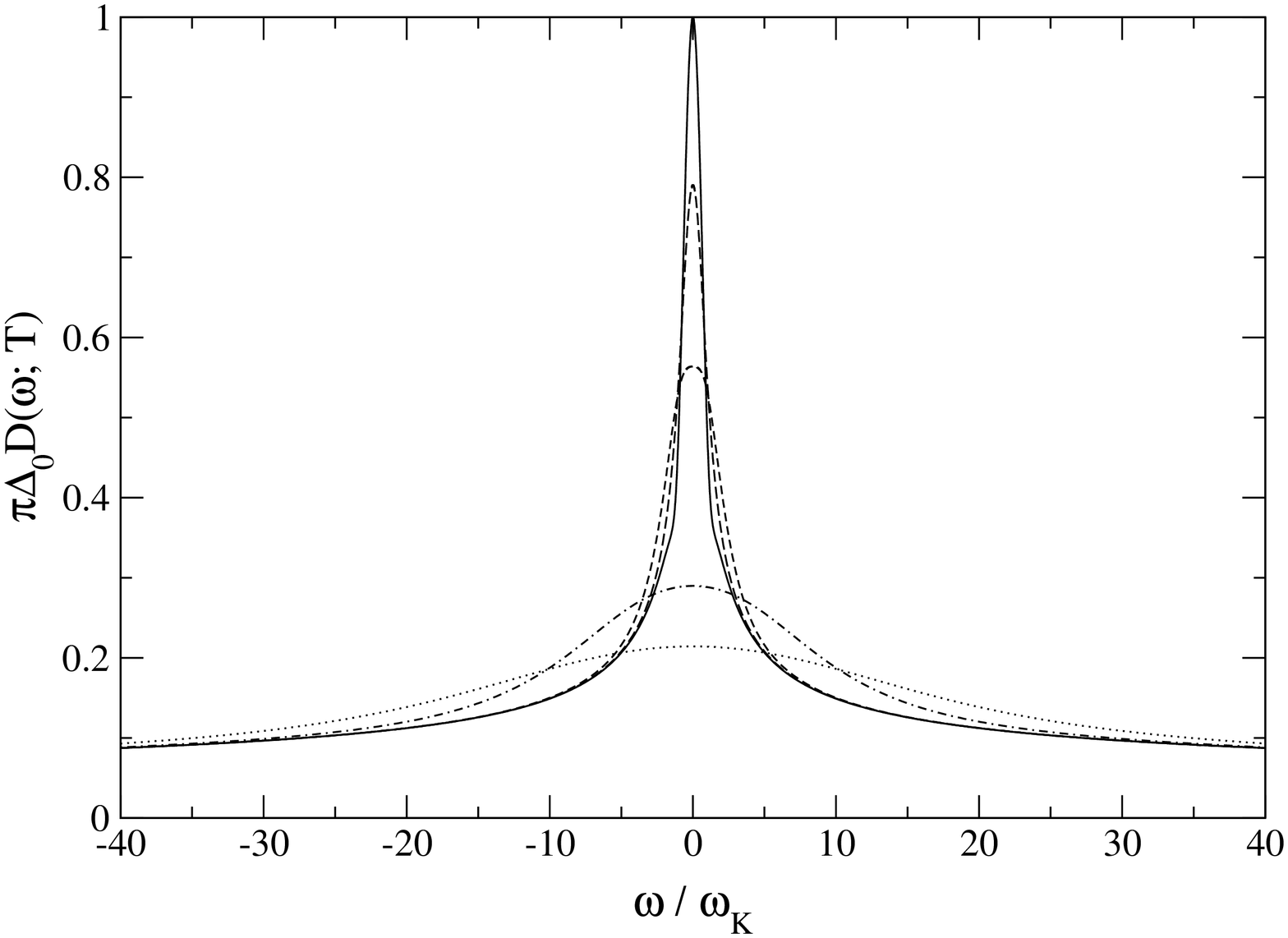,width=100mm,angle=270}
\vskip-5mm
\protect\caption{LMA(RPA) scaling spectra: $\pdod \ofwt$ vs $\w/\wk$ for $T/\wk = 0.1$ (solid line), 0.5 (long dash), 1 (short dash), 5 (point-dash) and 10 (dotted). LMA(RPA) denotes that $\cpm\ofw$ is given explicitly by the RPA-like ladder sum (see \S5.2).}
\label{fig:4}
\end{figure}

That such scaling indeed arises within the LMA upon progressively increasing $\ut$ is shown in figure~3 where, for fixed $T / \wk = 1$, $\pdod\ofwt$ is shown for three different interaction strengths, $\ut = 4$, $6$, $8$.  The inset shows the spectrum on an absolute scale vs $\w / \delno$, illustrating the exponential narrowing of the Kondo resonance with increasing $\ut$ ($\wk \propto \exp(-\pi U/8\delno)$). The main figure by contrast shows $\pdod \ofwt$ vs $\w / \wk$.  The spectra are indeed seen to approach asymptotically that of the Kondo scaling limit (obtained analytically in \S5, and also shown in figure~3); which behaviour is reached in practice for the $\ut$'s considered in figure~3.  The scaling illustrated above arises generically, and figure~4 shows the resultant LMA scaling spectra $\pdod \ofwt$ vs $\w / \wk$ for a range of different temperatures $T / \wk$.

Before proceeding to a detailed analysis of the scaling spectra, we comment briefly on the $T$-dependence of the transverse spin polarization propagator $\ppm \ofwpt$, whose spectral density $\cpm \ofwt$  enters the LMA self-energy $\sigs\ofwpt$ (\eqr{retsig1}).  Its $T$-dependence, arising from that of the $\pnpm\ofwpt$ bubble, is readily shown to occur on the scale $T \sim \delno$ and is non-negligible only for non-universal temperatures of this order (where it contributes \eg to the `spectral redistribution' discussed above in regard to figure~2).  The Kondo regime by contrast corresponds formally to finite $T / \wk$ in the limit $\wk \propto \exp(-\pi U / 8 \delno) \ra 0$.  Here the $T$-dependence of $\cpm \ofwt$ is thus entirely negligible (so from now on is referred to solely as $\cpm\ofw$); and the $T$-dependence of the LMA scaling spectrum is then controlled exclusively by $g(\w_1;\w_2)$ (\eqr{retsig2}) entering the LMA $\sigs\ofwpt$.

\section{Scaling Spectrum.}

Our aim now is to obtain analytically the $\w$- and $T$-dependence of the strong coupling Kondo scaling spectrum arising from the LMA; and, as in \cite{ref:5} where the $\teo$ limit was considered, we do so with only minimal assumptions about the form of the transverse spin spectrum $\cpm \ofw$.  In the following analysis the Kondo scale will appear in several guises, viz $\wm$, $\wmp$, $\wk$ and $\delno Z$ (with $Z$ the $\teo$ quasiparticle weight).  These are all however proportional to one another, and thus equivalent; reflecting the fact that in the Kondo limit the problem is characterized by a single low-energy scale.

Within the LMA the `primary' manifestation of the Kondo scale is $\wm$, the spin-flip scale determined via symmetry restoration (see \S2 and below), and embodied in a strong resonance in $\cpm \ofw$ centred on $\w = \wm$.  And the scaling spectrum is obtained by considering finite $\wt = \w / \wm$ and $\tt = T / \wm$ in the limit $\wm \propto \exp(-\pi U / 8 \delno) \ra 0$, which projects out the non-universal temperature and energy scales that are irrelevant to the scaling regime.  Hence, referring to equation~(2.3), the `bare' $\w = \wt \wm \equiv 0$ may be neglected; and likewise $\Delta (\w) = \Delta (\wt \wm)$ reduces to $\Delta(0) = -\rmi\delno$ (which applies to any metallic host, so that the following analysis is not confined to the wide-band AIM).  The scaling spectrum is thus given from equation~(2.3) by

\be
\label{eq:scalingd}
\pdod \ofwt = \frac{1}{2} \sum_\sigma \frac{( 1 + \delno\inv \sigsi \ofwpt )}{(\delno \inv \sigtsr \ofwpt )^2 + (1 + \delno \inv \sigsi \ofwpt )^2}
\ee

\noindent where $\sigts \ofwpt = - \sigma \frac{U}{2} + \sigs \ofwpt$ in the Kondo limit where the local moment saturates ($|\mu| \ra 1$).  The spectrum is naturally determined solely by the scaling behaviour of the self-energies, and the LMA $\sigu \ofwpt$ is given explicitly by (cf equation~(3.5)):

\bea
\label{eq:scsig}
\hspace{-20mm}\nonumber \sigu \ofwpt = U^2 \intall \frac{\rmd \w_1}{\pi}\intall \rmd \w_2\\
\cpm (\w_1)~ \frac{D_\da^0(\w_2)}{\w + \w_1 - \w_2 + \rmi 0^+}~\left[~\theta(\w_1) f(\w_2;T) + \theta(-\w_1)(1 - f(\w_2;T))\right].
\eea

In the strong coupling limit the transverse spin spectrum $\cpm \ofw$ has the following functional form \cite{ref:5,ref:6}

\numparts
\be
\label{eq:scpiform}
\frac{1}{\pi} \cpm \ofw = \frac{A}{\wm} F\ofwtil \theta \ofwtil,
\ee

\noindent naturally scaling in terms of $\wt = \w / \wm$; and

\be
\label{eq:scpiint}
\int\limits^\infty_0 \frac{\rmd \w}{\pi}~ \cpm \ofw = 1 = A\int\limits^\infty_0 \rmd y ~ F(y)
\ee
\endnumparts

\noindent which reflects the saturation of the local moment and complete suppression of double occupancy in the Kondo limit.  The function $F \ofwtil$ is distributed around and centred upon $\wt = 1$ (by definition of $\wm$), and $F \ofwtil \sim \wt$ as $\wt \ra 0$.  The above behaviour is readily shown to arise explicitly \cite{ref:6} with $\cpm \ofw$ given via the p-h ladder sum \eqr{pisum}, for which $F\ofwtil$ has the functional form

\be
\label{eq:pif}
F\ofwtil = \frac{\wt}{1 - 2 \alpha \wt + \wt^2}.
\ee

\noindent The specific form of $F\ofwtil$ is not however required in the following analysis (we shall need it only in \S5.2); and the important asymptotics of the scaling spectrum considered in \S5.1 are in fact independent of it.

From equations~(5.2,3), noting trivially that $f\ofwt = f(\wt;\tt)$, it follows directly that

\be
\label{eq:scsigi1}
\sigui \ofwpt = \pi U^2 A \int\limits^\infty_0 \rmd y ~ F(y) f(y+\wt;\tt) D_\da^0(\wm[y+\wt])
\ee

\noindent where $D_\da^0(\wm[y+\wt]) \equiv D_\da^0(0)$ since we consider finite $\wt$ with $\wm \ra 0$.  But in the strong coupling limit, $D_\da^0(0) = - \frac{1}{\pi} \im \sg_\da (0)$ is given (see \eqr{scriptg}) by

\be
\label{eq:scdo}
\pi U^2 D_\da^0(0) = 4 \delno
\ee

\noindent and hence:

\be
\label{eq:scsigi2}
\delno \inv \sigui \ofwpt = 4 A \int\limits^\infty_0 \rmd y ~ F(y) f(y+\wt;\tt).
\ee

\noindent As required in the Kondo limit this scales solely in terms of $\wt$ and $\tt$, with no $\ut$-dependence ($A$ being a pure number ${\cal O}(1)$ determined via \eqr{scpiint}).

From equations~(5.2,3), $\sigur \ofwpt$ is likewise given by

\be
\label{eq:scsigr1}
\sigur \ofwpt = A U^2 \intp \rmd y~F(y) \left\{ \intall \rmd \w_2~\frac{D_\da^0(\w_2)}{\wm[\wt + y] - \w_2}~f(\w_2;T) \right\}
\ee

\noindent (with a principle value implicit from now on).  For $\teo$, the $\w_2$-integration leads to $\re~\sg_\da^-(\wm[\wt + y])$, with $\re~\sg_\da^-\ofw$ logarithmically divergent as $\w \ra 0$.  To handle this at finite-$T$ we proceed as in \cite{ref:5}.  The lower limit of the $\w_2$-integration in \eqr{scsigr1} is replaced by a UV-cutoff, $-\lambda$, where $\lambda = \min[\frac{U}{2}, D]$ (with $D$ the host bandwidth); its precise value being immaterial in the following.  Rescaling $\w_2$ to $\wt_2 = \w_2 / \wm$ in \eqr{scsigr1} then leads to

\be
\label{eq:scsigr2}
\delno\inv \sigur \ofwpt = - \frac{4A}{\pi} \intp \rmd y~F(y) \int \limits^\infty_{-\tilde{\lambda}} \rmd \wt_2 ~ \frac{f(\wt_2;\tt)}{\wt_2 - [\wt + y]}
\ee

\noindent where \eqr{scdo} has again been used; and $\tilde{\lambda} = \lambda / \wm$ is thus defined, such that $\tilde{\lambda} \ra \infty$ in the Kondo limit.

For $\teo$, \eqr{scsigr2} reduces to \cite{ref:5}

\be
\label{eq:scsigr3}
\delno\inv \sigur \ofwpo = \frac{4}{\pi} \ln \left[ \frac{\lambda}{\wm} \right] - \frac{4A}{\pi} \intp \rmd y ~ F(y) \ln | y + \wt |
\ee

\noindent (where \eqr{scpiint} is used).  From this the explicit $U$-dependence of the strong coupling Kondo scale $\wm$ follows directly from symmetry restoration (\eqr{pinning}), viz $\sigur (0;0) = \half U$; namely

\numparts
\be
\label{eq:wmpdef1}
\wm = c \wmp
\ee

\noindent where

\be
\label{eq:wmpdef2}
\wmp = \lambda \exp (-\pi U / 8 \delno)
\ee

\noindent and $c$ a constant ${\cal O} (1)$ given by

\be
\label{eq:cint}
c = \exp \left[ -A \intp \rmd y ~ F(y) \ln y \right] .
\ee

\endnumparts

\noindent The prefactor to the exponent of $\wm$ or $\wmp$ is of course approximate (reflecting the UV-cutoff used above), but the exponent itself is exact.  And from \eqr{scsigr3}, $\sigtur \ofwpo \equiv \sigur \ofwpo - \sigur (0;0)$ at $\teo$ is given by:

\be
\label{eq:scsigro}
\delno \inv \sigtur \ofwpo = - \frac{4A}{\pi} \intp \rmd y~ F(y) \ln \left| 1 + \frac{\wt}{y}\right|
\ee

For the scaling spectrum $D\ofwt$ (\eqr{scalingd}) we finally require $\sigtur \ofwpt = - \frac{U}{2} + \sigur \ofwpt$, \ie

\numparts
\be
\label{eq:sigtr}
\sigtur \ofwpt = \left[ \sigur \ofwpt - \sigur \ofwpo \right] + \sigtur \ofwpo
\ee

\noindent with $\sigtur \ofwpo$ given above.  And using \eqr{scsigr2}, a straightforward calculation yields

\be
\label{eq:delsigrt}
\delno\inv \left[ \sigur \ofwpt - \sigur \ofwpo \right] = - \frac{4A}{\pi} \intp \rmd y ~ F(y) H\left(\frac{|y + \wt|}{\tt}\right)
\ee

\endnumparts

\noindent where

\be
\label{eq:hint}
H(x) = \intp \rmd \w ~ P\left( \frac{1}{\w + x} + \frac{1}{\w - x}  \right) \frac{1}{\exp\ofw + 1}
\ee

\noindent and the limit $\tilde{\lambda} \ra \infty$ has been taken with impunity.  As for $\delno \inv \sigui \ofwpt$ (\eqr{scsigi2}), equations~(5.12,13) show that $\delno \inv \sigtur \ofwpt$ also scales solely in terms of $\wt$ and $\tt$ with no $\ut$-dependence, as required from \eqr{scalingd} for the spectrum to scale thus.

Equations~(\ref{eq:scsigi2}) and (5.12,13), together with p-h symmetry (\eqr{phsigsym}), provide the basic results from which the $\wt$- and $\tt$-dependence of the scaling spectrum may be determined; we analyze them in the following sections.

Before proceeding we note that the function $H(x)$ (\eqr{hint}) is given in closed form by

\be
\label{h}
H(x) = -\ln \left[ \frac{|x|}{2\pi}\right] + \re~\psi\left (\frac{1}{2} + \rmi \frac{|x|}{2\pi}\right)
\ee

\noindent with $\psi (z) = \rmd \ln \Gamma (z) / \rmd z$ the digamma function; and is readily shown to be of form

\numparts
\be
\label{eq:hform}
H(x) = - \ln \left[ \frac{|x| / \pi}{\left( 1 + \left[ \frac{|x|}{\pi} \right]^2\right)^\half} \right] - h(x)
\ee

\noindent where $h(x)$ is non-singular.  A useful approximation to $h(x)$ is

\be
\label{eq:happrox}
h(x) \simeq \frac{\gamma}{1 + \frac{3}{2}\gamma \left[ \frac{x}{\pi} \right]^2}
\ee
\endnumparts

\noindent where $\gamma = {\mathbb C} + \ln 2$ and ${\mathbb C}$ ($=0.577...$) is Euler's constant.  As required in the following section, this ensures that the leading asymptotics of $H(x)$ as $x \ra 0$ and $x \ra \infty$ are captured exactly; being given by

\numparts

\be
\label{eq:holim}
H(x)~ {^{x \ra 0}_{~\sim}}~ -\ln \left[ \frac{|x|}{\pi} \right] - \gamma + {\cal O}(x^2)
\ee

\noindent and

\be
\label{eq:hinflim}
H(x)~ {^{x \ra \infty}_{~\sim}}~ - \frac{1}{6} \left[ \frac{\pi}{x} \right] ^2 + {\cal O}(x^{-4})
\ee

\endnumparts

\noindent (which is essentially a Sommerfeld expansion).

\subsection{Spectral asymptotics.}

We now consider the asymptotic behaviour of the LMA scaling spectrum and, relatedly, the corresponding single self-energy $\Sigma\ofwpt$.  This encompasses both the spectral tails, $|\wt| \gg 1$ for any $\tt$, as well as high temperatures, $\tt \gg |1 + \wt|$ for any $\wt$; and is not dependent on details of the function $F(\wt)$ (equation~(5.3)) that determines the spectral density of transverse spin excitations.  In practice the specific form of $F(\wt)$ is required only for low $\wt$ and $\tt$, as considered in \S5.2.

From \eqr{scsigi2}, using \eqr{scpiint} and recalling that $F\ofwtil$ is distributed around and centered on $\wt = 1$, $\delno \inv \sigui \ofwpt$ is given for $|\wt| \gg 1$ and any $\tt$ by

\be
\label{eq:lwsigi}
\delno \inv \sigui \ofwpt \sim 4 f(\wt;\tt).
\ee

\noindent This holds also for $\tt \gg |1+\wt|$ and all $\wt$.  Turning now to equation~(5.13) for $\sigtur \ofwpt$, note first from \eqr{scsigro} that for $\teo$, $\delno \inv \sigtur \ofwpo \sim -\frac{4A}{\pi} \int^\infty_0 \rmd y~F(y) \ln |\wt/y|$ for $|\wt| \gg 1$; and hence, using equations~(5.11$a$,$c$) and (\ref{eq:scpiint}), 

\be
\label{eq:lwsigr}
\delno \inv \sigtur \ofwpo \sim -\frac{4}{\pi} \ln|\wp|
\ee

\noindent with

\numparts

\be
\label{eq:wpdef}
\wp = \w / \wmp
\ee

\noindent thus defined.  \Eqr{delsigrt} for $|\wt|\gg 1$ likewise reduces (again using \eqr{scpiint}) to $\delno \inv [\sigur \ofwpt - \sigur \ofwpo] \sim - \frac{4}{\pi}H(|\wt|/\tt) = - \frac{4}{\pi}H(|\wp|/\tp)$, where

\be
\label{eq:tpdef}
\tp = T / \wmp
\ee

\endnumparts

\noindent is correspondingly defined.  Combined with \eqr{lwsigr}, and using \eqr{hform} for $H(x)$, equation~(5.13) for $\sigtur \ofwpt$ thus reduces for $|\wt| \gg 1$ to

\be
\label{eq:lwsigr2}
\delno \inv \sigtur \ofwpt \sim -\frac{4}{\pi} \left( \ln \left[ \left([\pi\tp]^2+ |\wp|^2\right)^\half \right] - h \left( \frac{|\wp|}{\tp}\right) \right);
\ee

\noindent which is readily shown to hold also for $\tt \gg |1+\wt|$ and all $\wt$.

The asymptotic behaviour of the Kondo limit scaling spectrum then follows from \eqr{scalingd}, using \eqr{lwsigi} (and p-h symmetry), as

\bea
\label{eq:lwscd}
\hspace{-25mm}\pdod \ofwt \sim \\
\vspace{4mm}
\nonumber\hspace{-10mm}\frac{1}{2}\left\{ \frac{[1+4f(\wp;\tp)]}{[\delno \inv \sigtur \ofwpt]^2\! +\! [1\! +\! 4 f(\wp;\tp)]^2} + \frac{[1+4f(-\wp;\tp)]}{[\delno \inv \sigtur \ofwpt]^2\! +\! [1\! +\! 4 f(-\wp;\tp)]^2}   \right\}
\eea

\noindent with $\delno \inv \sigtur \ofwpt$ from \eqr{lwsigr2}.  Three points should be noted here:  

\noindent(i) For $|\wp| \gg \max(1, \tp)$ (where $f(\wp;\tp) \sim \theta(-\wp)$), \eqr{lwsigr2} for $\sigtur \ofwpt$ reduces (using \eqr{happrox}) to its $\teo$ limit, \eqr{lwsigr}; hence the spectrum for $|\wp| \gg \max(1,\tp)$ reduces to

\be
\label{eq:tails}
\pdod \ofwt \sim \frac{1}{2} \left\{ \frac{1}{\left[ \frac{4}{\pi} \ln |\wp| \right]^2 +1} + \frac{5}{\left[ \frac{4}{\pi} \ln |\wp| \right] ^2 +25}    \right\} .
\ee

\noindent This is also the result of \cite{ref:5} for the $\teo$ spectral `tails', there shown to dominate the $\teo$ scaling spectrum (down to $\w \sim \wk$, the crossover to Fermi liquid behaviour occurring only on the lowest energy scales $|\wp| \ll 1$); and to be quantitatively accurate in comparison to NRG calculations for $|\wp| \gtrsim 5$.  The arguments leading to \eqr{tails} show that the same behaviour arises also at finite-$T$, for sufficiently high frequencies $|\wp| \gg \max(1,\tp)$.  This `common tail' behaviour is indeed seen in the scaling spectra shown in figure~4; and is evident also in recent finite-$T$ NRG calculations for the Kondo model itself (\cite{ref:21}, figure~1 therein).

\noindent(ii) For $\tp \gg |1+\wp|$ by contrast (where $f(\wp;\tp) \sim \half$), \eqr{lwsigr2} for $\sigtur\ofwpt$ reduces (again using \eqr{happrox}) to

\be
\label{eq:swotsigr}
\delno \inv \sigtur \ofwpt \sim -\frac{4}{\pi} \ln [\epsilon \tp]
\ee

\noindent where $\epsilon = \frac{\pi}{2}\exp(-{\mathbb C})$; and the spectrum is thus given by

\be
\label{eq:swotd}
\pdod \ofwt \sim \frac{1}{3 + \frac{16}{3\pi^2}\ln^2 [\epsilon\tp]}.
\ee

\noindent For $\tp \gg |1+\wp|$ the $\w$-dependence of $D\ofwt$ is thus essentially irrelevant; as is seen in figure~4 for \eg $T / \wk = 10$, where the spectrum flattens out at low frequencies.  \Eqr{swotd} gives in particular the asymptotic high-temperature behaviour of the Fermi level spectrum $D\ofot$, 
which will be considered further in \S5.2

\noindent (iii)  For $\tp \gg 1$, the leading asymptotic behaviour of 
$D\ofwt$ 
is clearly dominated by the logarithmic growth of $\sigtur \ofwpt$ (since $f(\wp;\tp) \leq 1$); and is given from \eqr{lwscd} by

\bea
\nonumber\pdod \ofwt &\sim \frac{3}{[\delno \inv \sigtur \ofwpt]^2}\\
\label{eq:lwallt}
&= \frac{3\pi^2}{16} \frac{1}{\ln^2\left[ \sqrt{(\pi\tp)^2 + |\wp|^2}~\rme^{-h\left(\frac{|\wp|}{\tp}\right)} \right]}
\eea

\noindent which form will prove important in \S6 where the resistivity $\rho(T)$ is considered.

Finally, the corresponding asymptotics of the conventional single self-energy $\Sigma \ofwpt = \sigr \ofwpt - \rmi \sigi \ofwpt$ may also be obtained.  $\Sigma \ofwpt$ is related generally to the LMA self-energies $\sigts \ofwpt$ by \eqr{fullsig}; with the $U=0$ propagator therein reducing trivially to $g\ofw \equiv g(0)  = -\rmi/\delno$ in the Kondo limit.  Using equations~(5.18,21) the leading asymptotic behaviour of $\Sigma\ofwpt$ (applicable for $|\wp| \gg 1$ and all $\tp$, or $\tp \gg |1+\wp|$ and all $\wp$) is thereby found to be:

\numparts

\be
\label{eq:fsigr}
\delno \inv \sigr \ofwpt \sim  -\sgn\ofw \tanh \left( \frac{|\wp|}{2\tp}\right) \frac{16}{3\pi} \ln \left[\left([\pi\tp]^2 + |\wp|^2\right)^\half \right]
\ee

\be
\label{eq:fsigi}
\delno \inv \sigi \ofwpt \sim  \frac{16}{3\pi^2} \ln^2 \left[ \left([\pi\tp]^2 + |\wp|^2 \right)^\half  \right].
\ee

\endnumparts

For $|\wp| \gg \max(1, \tp)$, the $T$-dependence of $\Sigma \ofwpt$ is irrelevant, and equations~(5.27) reduce to $\delno \inv \sigr \ofwpt \sim - \sgn \ofw \frac{16}{3\pi} \ln |\wp|$ and $\delno \sigi \ofwpt \sim \frac{16}{3\pi^2} \ln^2|\wp|$.  Again, these coincide with the $\teo$ result, also found in \cite{ref:5} to give good agreement with $\teo$ NRG calculations.  For $\tp \gg |1+\wp|$ by contrast $\tp$ rather than $\wp$ controls the logarithms, which incipient divergence then dominates the imaginary part of the single self-energy, $\sigi \ofwpt \sim \ln ^2 (\tp)$.

\subsection{Scaling spectra: all scales.}

The asymptotic behaviour obtained above encompasses high frequencies for all temperatures, as well as high temperatures for any frequency; and is independent of the detailed form of the transverse spin spectrum $\cpm \ofw$.  To obtain the LMA scaling spectrum on all energy/temperature scales, and in particular the low-$(\w,T)$ behaviour, requires by contrast a full specification of $\cpm \ofw$.  This we now turn to, considering as in \cite{ref:5} two related variants of the function $F(\wt)$ that determines $\cpm\ofw$ via equation~(5.3).

In the first, referred to from now on as the LMA(RPA), $\cpm \ofw$ is obtained explicitly from the RPA-like p-h ladder sum \eqr{pisum}.  $F(\wt)$ then has the form \eqr{pif} \cite{ref:6}, and it is known \cite{ref:5,ref:6} that in the strong coupling limit $\ut \gg 1$, $\alpha \ra 1$ and $A \ra 0$ such that  $Af(y) = \delta(y-1)$.  From equation~(5.3$a$), $\frac{1}{\pi} \cpm\ofw = \delta (\w - \wm)$ thus reduces to a simple $\delta$-function centred on $\wm$.  The Kondo limit scaling spectrum arising from the LMA(RPA) then follows in closed form from equations~(5.1,7,12,13), together with equation~(5.15) for $H(x)$.  It indeed recovers precisely the scaling spectra obtained numerically in \S4 by progressively increasing $\ut$, and illustrated in figure~4.

The limitations of the above LMA(RPA) reside in the $\delta$-function form of $\cpm \ofw$ -- in reality $\cpm \ofw$ will have non-zero width, reflecting a finite lifetime for the spin-flip excitations.  To encompass this we proceed as in \cite{ref:5}, retaining the functional form \eqr{pif} for $F(\wt)$ (which has finite width provided $\alpha \ne 1$), and employing a high frequency cutoff $\wct$ to render $F(\wt)$  normalizable (\eqr{scpiint}).  The width parameter $\alpha$ is then determined by requiring that the leading low-frequency behaviour of the imaginary part of the $\teo$ single self-energy is recovered exactly.  As explained in \cite{ref:5} this requires $A = \half [ \wm / \delno Z]^2$ (with $Z$ the $\teo$ quasiparticle weight), which in turn determines $\alpha$ uniquely for a chosen cutoff $\wct$.  The latter is of course arbitrary but, as one expects, results are not sensitive to it \cite{ref:5}; in practice, as in \cite{ref:5}, we choose $\wct = 10$.  From now on we refer to this simply as the LMA, and recall in passing that \cite{ref:5} the ($\teo$) HWHM Kondo scale $\wk = 0.691 \wmp$ for both it and the LMA(RPA).

\begin{figure}
\centering\epsfig{file=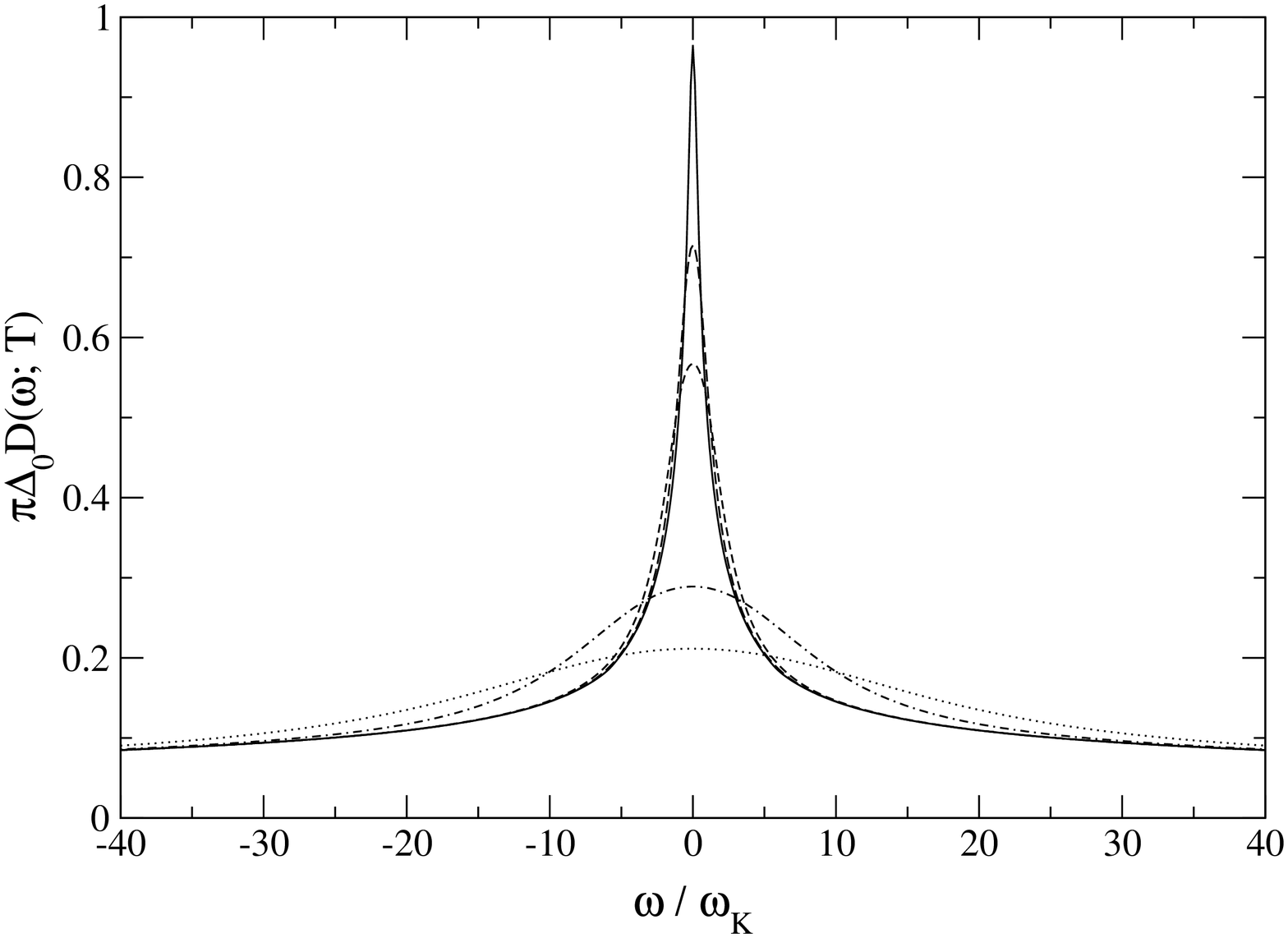,width=100mm,angle=270}
\vskip-5mm
\protect\caption{LMA scaling spectra obtained as explained in text: $\pdod \ofwt$ vs $\w / \wk$ for $T/\wk$ = 0.1 (solid line), 0.5 (long dash), 1 (short dash), 5 (point-dash) and 10 (dotted).}
\label{fig:5}
\end{figure}

The resultant LMA scaling spectra are shown in figure~5 for the same temperatures considered in figure~4 for their LMA(RPA) counterparts, $T / \wk = 0.1$, $0.5$, $1$, $5$ and $10$.  The differences between the two are seen to be rather minor; in fact for $T/\wk \gtrsim 1$ the corresponding spectra are essentially coincident for all frequencies.  Not unexpectedly, the primary differences arise at low temperatures and frequencies, comparison of figures~4,5 showing that for $T / \wk < 1$ the Kondo resonance diminishes less rapidly with increasing temperature in the LMA(RPA).

To pursue this we focus on the thermal evolution of the spectrum at the Fermi level, $D(\w=0;T)$; which is related to the single self-energy by

\numparts

\be
\label{eq:dzero}
\pdod \ofot = [1 + \delno \inv \sigi \ofot ] \inv
\ee

\noindent with

\be
\label{eq:sigio}
\sigi \ofot = \sigui \ofot + \frac{\delno \inv [\sigtur \ofot ]^2}{1 + \delno \inv \sigui \ofot}
\ee

\endnumparts

\noindent given in terms of the LMA self-energies (as follows generally from \eqr{fullsig} using p-h symmetry).  The resultant $T/\wk$-dependence of $\pdod \ofot$ is shown in figure~6, from which the LMA and LMA(RPA) are indeed seen to be essentially coincident for $T / \wk \gtrsim 1$, while differing quite significantly at lower temperatures.  For the LMA(RPA) it is readily shown using the preceding results that the leading low-$T$ dependence is $\sigi \ofot \sim {\cal O}([T / \wk]^4)$, and hence $\pdod \ofot \sim 1 - {\cal O}([ T / \wk]^4)$.  For the LMA by contrast, the corresponding $T$-dependence is of the correct form \cite{ref:2} $\sigi \ofot \sim {\cal O}([T / \wk ]^2)$.  The leading low $\tt$ ($=T / \wm$) and $\wt$ ($= \w / \wm$) dependence of the LMA $\sigi\ofwpt$ can in fact be obtained analytically using equations~(5.7,12,13) with equation~(5.4) for $F(\wt)$, together with equation~(2.13) relating $\Sigma \ofwpt$ to the LMA $\sigts \ofwpt$.  The resultant $\sigi \ofwpt$ is thereby found to be

\numparts

\bea
\label{eq:fs1}
\delno \inv \sigi \ofwpt & \sim A \left[ \wt^2 + \frac{\pi^2}{3}\tt^2 \right] \\
\label{eq:fs2}
&= \frac{1}{2} \left[ \left( \frac{\w}{\delno Z} \right)^2 + \frac{\pi^2}{3} \left( \frac{T}{\delno Z}\right)^2 \right]
\eea

\endnumparts

\noindent where $A = \half [ \wm / \delno Z]^2$ is used, and hence $\pdod \ofot \sim 1 - \frac{\pi^2}{6} [ T / \delno Z]^2$.  Equation~(5.29) differs from the known exact result \cite{ref:2,ref:22} only in the coefficient of $[T / \delno Z]^2$, the exact value for which is $\frac{\pi^2}{2}$.

\begin{figure}
\centering\epsfig{file=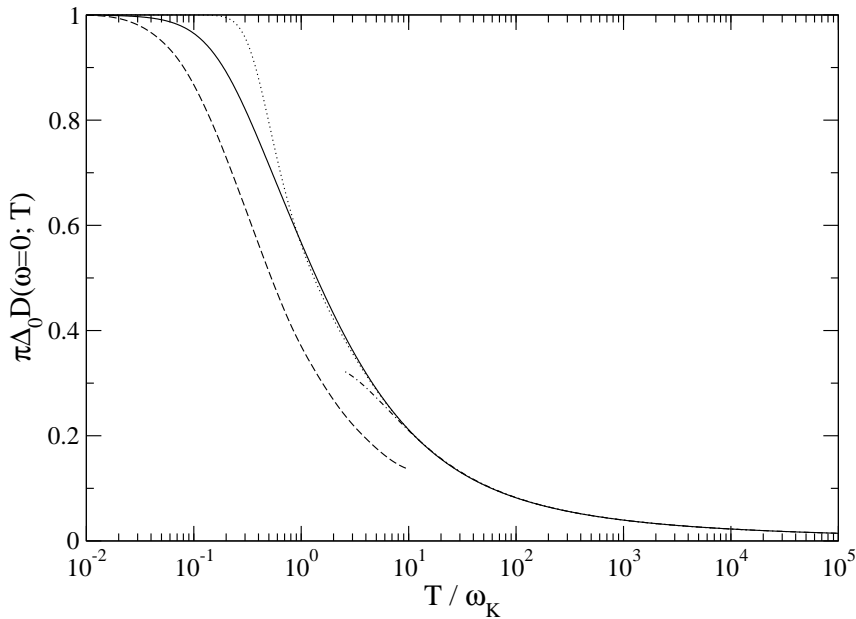,width=100mm,angle=270}
\vskip-5mm
\protect\caption{Fermi level spectrum $\pdod(\w\!=\!0;T)$ vs $T/\wk$ for the LMA (solid line) and LMA(RPA) (dotted); the point-dash line shows the high temperature asymptotic behaviour \eqr{swotd}.  NRG results \cite{ref:21} for the Kondo model are also shown (dashed line).}
\label{fig:6}
\end{figure}

Figure~6 also shows the explicit high temperature asymptotic result for $\pdod \ofot$, \eqr{swotd}, which is essentially perfect quantitatively for $T / \wk \gtrsim 10$, and accurate to within 10\% down to $T / \wk \simeq 5$.  We believe in fact that the leading high-temperature form from \eqr{swotd}, $\pdod \ofwt \sim 3 \pi^2 / [16 \ln^2 (T / \wk)]$ is exact; for essentially the same reasons that the present theory captures the exact high temperature behaviour of the resistivity, as discussed in the following section.

Finally, the LMA results for $\pdod \ofot$ are also compared in figure~6 to recent NRG calculations \cite{ref:21} performed directly on the Kondo model (with $\rho_{\rm host}(0)J = 0.19$); the NRG results for $D\ofot / D(0;0)$ are shown, and allow for the slight offset in the NRG $\delno D (0;0)$ below its exact value of $1/\pi$.  Agreement is qualitatively good although, as expected from equation~(5.29$b$) where the coefficient of the leading low-$T$ term is $\frac{1}{3}$ of its exact value, the LMA  $\pdod \ofot$ is higher than the NRG data at low-temperatures; and this behaviour persists over the temperature range for which NRG data has been published.  No great leap of faith is however required to anticipate that NRG results for higher temperatures are likely to concur with the high-temperature behaviour of the present theory.

\section{Resistivity.}

One reason why a knowledge of the single-particle spectrum $D\ofwt$ is important, is that transport properties may be obtained directly from it, as mentioned in \S2.  We consider here the resistivity $\rho\oft$ (\eqr{rhodef}), focussing naturally on the Kondo scaling regime; and for which the present theory is found to be asymptotically exact at high temperatures, while also recovering the correct Fermi liquid form $\rho\oft / \rho \ofo \sim 1 - \beta [ T/\wk]^2$ as $T \ra 0$ \cite{ref:19}, and yielding rather good agreement with NRG calculations \cite{ref:16}.

\begin{figure}[t]
\centering\epsfig{file=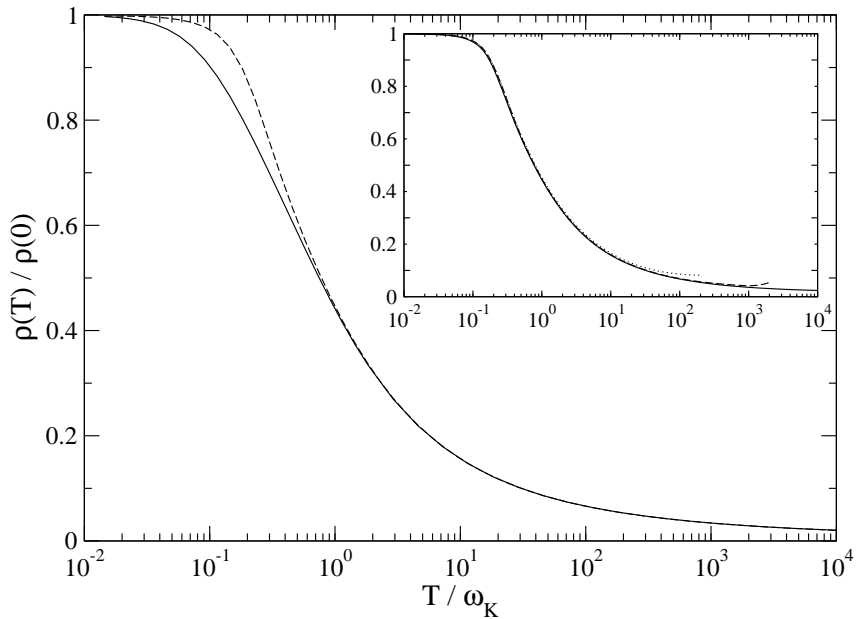,width=100mm,angle=270}
\vskip-5mm
\protect\caption{Resistivity $\rho\oft / \rho\ofo$ vs $T/\wk$ for the Kondo scaling limit; LMA (solid line), LMA(RPA) (dashed).  {\bf Inset}:  Approach to universal scaling with increasing $\ut$, $\rho\oft / \rho\ofo$ vs $T/\wk$ for the LMA(RPA) with $\ut = 4$ (dotted line), 6 (dashed) and 8 (solid line).}
\label{fig:7}
\end{figure}

First, we consider briefly the approach to the Kondo scaling limit with increasing interaction strength $\ut$.  The inset to figure~7 shows the resultant $\rho\oft / \rho \ofo$ vs $T / \wk$ obtained from the LMA(RPA), for three different $\ut$: 4, 6  and 8.  From this the rapid approach to the limiting scaling form is self-evident.  Deviations from scaling behaviour naturally set in for non-universal temperatures $T$ on the order of $\delno$.  This leads to the characteristic $\ut$-dependent `upturn' in $\rho\oft$ evident in figure~7 (inset); which with increasing $\ut$ moves rapidly to exponentially large vales of $T / \wk$ ($\wk \propto \exp(-\pi U / 8 \delno)$), such that more and more of the Kondo scaling form is recovered.  The main part of figure~7 shows the $T / \wk$-dependence of the resistivity in the Kondo limit, obtained using the analytical results of \S5.  Results for both the LMA(RPA) and LMA are shown and, as expected from the corresponding discussion of spectral asymptotics (\S5.1), are coincident in practice for $T/\wk \gtrsim 1$.

We thus consider first the high temperature behaviour of $\rho\oft$.  This is controlled by the high temperature form of $D\ofwt$, the leading asymptotic behaviour of which is given explicitly by \eqr{lwallt}.  Inserting the latter into \eqr{rhodef} for $\rho\oft$, and transforming the integration variable therein from $\w$ to $x = \w / T \equiv \wp / \tp$, yields straightforwardly the leading asymptotic behaviour for $T / \wk \gg 1$,

\be
\label{eq:asymprho}
\frac{\rho\oft}{\rho\ofo} \sim \frac{3\pi^2}{16}\frac{1}{\ln^2(T / \wk)}
\ee

\noindent (where $\wk \propto \wmp$ is used).  This is indeed the exact high-temperature asymptote for the $S=\half$ Kondo/s-d model, first obtained as the leading logarithmic sum of parquet diagrams by Abrikosov \cite{ref:17}.  Note also that this behaviour mirrors closely the leading high-{\em frequency} asymptotics of the $\teo$ single-particle spectrum, given from \eqr{tails} by $\pdod \ofwo \sim 3 \pi^2/[16\ln^2(|\w|/\wk)]$.

Resummation of parquet diagrams for the Kondo/s-d model leads further to the well known Hamann approximation for $\rho\oft$ \cite{ref:18},

\be
\label{eq:hamann}
\frac{\rho\oft}{\rho\ofo} \simeq \frac{1}{2}\left\{ 1 - \frac{\ln(T / \tkh)}{\left[\ln^2 (T / \tkh) + \frac{3\pi^2}{4}\right]^\half} \right\}
\ee

\begin{figure}{t}
\centering\epsfig{file=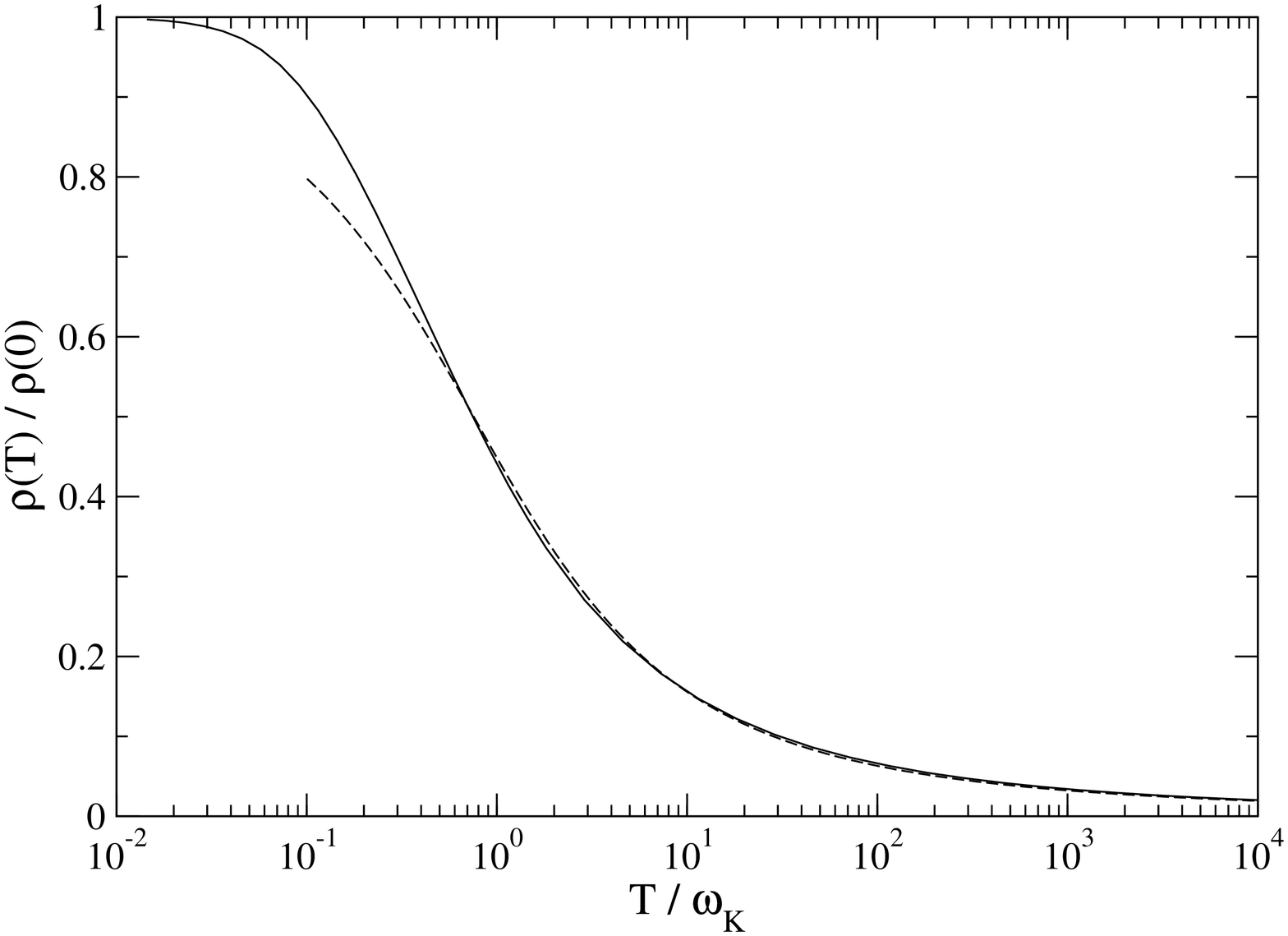,width=100mm,angle=270}
\vskip-5mm
\protect\caption{$\rho\oft / \rho\ofo$ vs $T/\wk$ for the LMA (solid line) compared to the Hamann approximation \eqr{hamann} (dashed line).}
\label{fig:8}
\end{figure}

\noindent which correctly recovers \eqr{asymprho} to leading order for $T / \wk \gg 1$ ($\tkh \propto \wk$).  A fit of \eqr{hamann} to our results is shown in figure~8 (with $\tkh / \wk = 0.76$).  From this, as found also in NRG calculations (\cite{ref:16}, see also figure~9 below), it is seen that the Hamann result accounts well for $\rho\oft$ down to temperatures $T / \wk \sim 1$; while the leading high-temperature behaviour \eqr{asymprho} is by contrast accurate only for $T / \wk \gtrsim 10^2$.  We note too the parallel between the Hamann results for $\rho\oft$, and \eqr{tails} for the high-frequency behaviour of the single-particle spectrum for $\teo$; where the latter in practice agrees well with NRG results for $D\ofwo$ down to $\w / \wk\sim 1$ (\cite{ref:15}, see \eg figure~2 therein), even though its leading high-frequency asymptote $\propto 1 / \ln^2 (|\w| / \wk)$ is accurate only for $\w / \wk \gtrsim 10^2$.

We turn now to the low-temperature asymptotics of $\rho \oft$, which as first shown by Nozi$\grave{\rm e}$res \cite{ref:19} has the characteristic Fermi liquid form

\be
\label{eq:rfermi}
\frac{\rho\oft}{\rho\ofo} \sim 1 - b\left[\frac{T}{\delno Z}\right]^2
\ee

\noindent with the exact coefficient $b = \pi^2$; and where $\delno Z$ (with $Z$ the $\teo$ quasiparticle weight) is related to the conventional Kondo temperature $\tk$, defined such that the static impurity susceptibility $\chi_{\rm imp} = (\gmb)^2/4\tk$, by $\tk = \frac{\pi}{4}\delno Z$.  The $T^2$-expansion of $\rho\oft$ may be extended to higher order via the methods of boundary conformal field theory \cite{ref:23}, and in principle by the more generally applicable renormalized perturbation theory \cite{ref:24,ref:25}.  The former has been carried out exactly up to $(T / \delno Z)^6$ \cite{ref:23}; but this impressive feat barely extends the applicability of the $T^2$-expansion beyond $T / \delno Z \sim 10\inv$, up to which the Fermi liquid result \eqr{rfermi} is accurate in practice.

As shown in \cite{ref:16}, the leading low-$T$ behaviour of $\rho\oft$ for the AIM generally is related to the low-$(\w,T)$ behaviour of the single-particle spectrum by

\be
\label{eq:aimrfl}
\frac{\rho\oft}{\rho\ofo} \sim \left\{ 1 + \frac{(\pi T)^2}{6} \left( \frac{\partial^2 \sa\ofwo}{\partial\w ^2}\right)_{\w = 0} + \frac{T^2}{2} \left(\frac{\partial^2 \sa\ofot}{\partial T ^2}\right)_{T = 0}   \right\}
\ee

\noindent where $\sa\ofwt = \pdod \ofwt$.  In the Kondo limit the exact results are \cite{ref:2,ref:16} $\sa \ofwo \sim 1 - \frac{3}{2}[\w / \delno Z]^2$ and $\sa \ofot \sim 1 - \frac{\pi^2}{2}[T / \delno Z]^2$; \eqr{rfermi} thus follows.  The leading low-$T$ behaviour of $\rho\oft$ from the present local moment approach can likewise be determined.  It recovers the Fermi liquid form \eqr{rfermi}, scaling in terms of $T/\delno Z$, although the coefficient $b$ is not exact.  For the LMA, $\sa\ofwo$ to leading order (and hence the first non-trivial term in \eqr{aimrfl}) is exact, while $\sa \ofot = [1 + \delno \inv \sigi \ofot]\inv \sim 1 - \frac{\pi^2}{6}[T / \delno Z]^2$ (from equation~(\ref{eq:fs2}));  so the coefficient $b = 2\pi^2/3$.  And for the LMA(RPA), where $\sa \ofwo \sim 1 - [\w / \delno Z]^2$ \cite{ref:5,ref:6} and $\sa \ofot \sim 1 - {\cal O}([T/\delno Z]^4)$ (\S5.2), $b = \pi^2/3$.  The difference between the LMA(RPA) and LMA at low temperatures is of course seen directly in figure~7.

\begin{figure}
\centering\epsfig{file=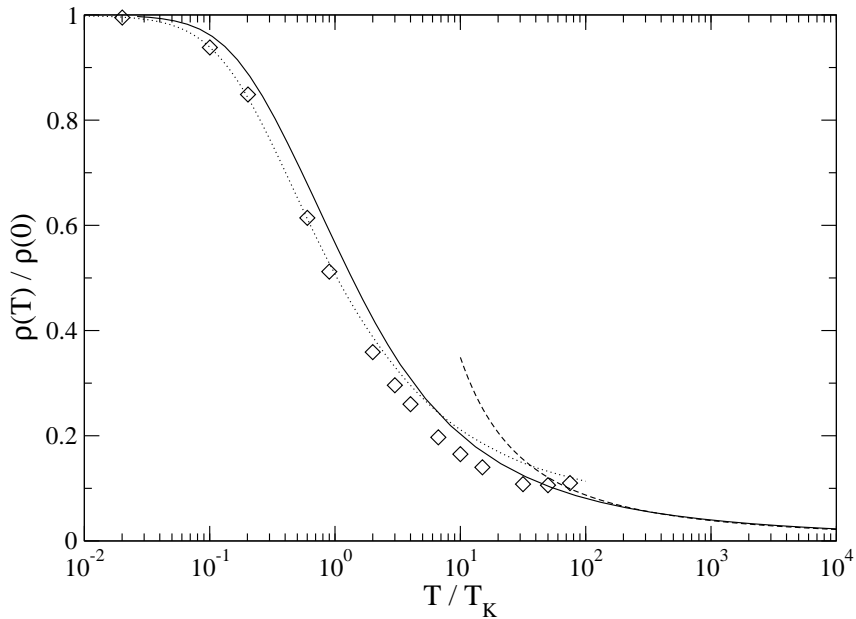,width=100mm,angle=270}
\vskip-5mm
\protect\caption{$\rho\oft / \rho\ofo$ vs $T/\tk$ (with $\tk$ defined such that $\chi_{\rm imp} = (\gmb)^2/4\tk$).  LMA (solid line) and NRG results \cite{ref:16} for the AIM with $\ut = 4$ (diamonds).  The exact high-$T$ asymptote $\rho\oft / \rho\ofo \sim 3\pi^2/[16\ln^2(T/\tk)]$ \cite{ref:17} is also shown (dashed line); as are results from a recent low-T approximation \cite{ref:26} exploiting integrability (dotted line).  See text for discussion.}
\label{fig:9}
\end{figure}

In figure~9 the LMA $\rho\oft / \rho \ofo$ vs $T / \tk$ is compared directly to the NRG results of Costi, Hewson and Zlati$\acute{\rm c}$ \cite{ref:16}.  The latter were obtained for the AIM with $\ut = 4$; and the final two/three high-T NRG  points shown correspond  to $T \sim {\cal O}(\delno)$, thus exhibiting the characteristic non-universal upturn from Kondo scaling behaviour mentioned above in relation to figure~7 (inset).  We regard the agreement with NRG as rather good, the more so given the relative simplicity of the local moment approach and its ready extendibility to a much wider range of problems than the integrable AIM.  Figure~9 also shows results from a recent approach \cite{ref:26} that, while unable to access single-particle dynamics, focusses directly on transport (strictly the linear differential conductance, which is however known to differ negligibly from $\rho\oft$ \cite{ref:21}).  This method exploits the integrability of the AIM via an approximate treatment of the Bethe ansatz equations, and is designed to capture the low-$T$ behaviour.  That is does so very well up to $T/\tk \sim 1$ is evident from figure~9; although it does not appear to recover the high-temperature Abrikosov asymptote \eqr{asymprho}  (referred to in \cite{ref:26} as the 1-loop RG result after its recent essential rediscovery in \cite{ref:27} via an elegant RG approach).  The latter (specifically $3\pi^2 / [16\ln^2(T/\tk)]$) is also shown in figure~9 and is recovered by the present work, in practice for $T/\tk \gtrsim 10^2$.

\section{Conclusion.}

The subject of this paper has been single-particle dynamics of the Anderson impurity model, pursued via the local moment approach extended to finite temperature, and with a natural emphasis on the strongly correlated Kondo scaling regime.  The approach yields thereby a rich, and rather successful description of the thermal destruction of the Kondo resonance and hence of transport properties such as the resistivity.  This augurs well for the future, since the LMA is neither technically difficult nor confined to the integrable, metallic AIM considered here; suggesting its practical viability as a potentially powerful route to dynamics and transport properties of a wider range of quantum impurity models, as well as \eg the periodic Anderson and Hubbard \cite{ref:28} models within the framework of dynamical mean field theory.  Problems of this ilk are currently under study, and will be considered in subsequent publications.

\ack
We are grateful to the EPSRC, the Leverhulme Trust and the British Council for financial support.

\renewcommand{\appendix}{%
	\setcounter{section}{1}
	\setcounter{equation}{0}
	\renewcommand{\thesection}{\Alph{section}}
	\renewcommand{\theequation}{\Alph{section}.\arabic{equation}}
}

\vskip+5mm

\end{document}